# Better Than Their Reputation? On the Reliability of Relevance Assessments with Students

Philipp Schaer

GESIS – Leibniz Institute for the Social Sciences,
Unter Sachsenhausen 6-8, 50667 Cologne, Germany
`philipp.schaer@gesis.org`

**Abstract.** During the last three years we conducted several information retrieval evaluation series with more than 180 LIS students who made relevance assessments on the outcomes of three specific retrieval services. In this study we do not focus on the retrieval performance of our system but on the relevance assessments and the inter-assessor reliability. To quantify the agreement we apply Fleiss' Kappa and Krippendorff's Alpha. When we compare these two statistical measures on average Kappa values were 0.37 and Alpha values 0.15. We use the two agreement measures to drop too unreliable assessments from our data set. When computing the differences between the unfiltered and the filtered data set we see a root mean square error between 0.02 and 0.12. We see this as a clear indicator that disagreement affects the reliability of retrieval evaluations. We suggest not to work with unfiltered results or to clearly document the disagreement rates.

## 1   Introduction

During the last three years we conducted several information retrieval evaluation series regarding different retrieval-supporting services. More than 180 LIS students made relevance assessments on the outcomes of three specific retrieval services. These three services were designed to compensate typical problems that arise in metadata-driven digital libraries, which are not adequately handled by a simple TF*IDF based retrieval. The services are: a co-word analysis based query expansion mechanism and re-ranking via Bradfordizing and Author Centrality. The overall system and the value added services, outlined in section 3.1, are well documented and previous evaluation results were presented at conferences and journals before [14, 15, 18].

While we did a study on the inter-assessor or inter-rater agreement for the first year of our evaluation, until now we did no meta-analysis regarding the time period of three years or the effects of the large number of students. Therefore the questions for this work are: (1) How good and reliable are the relevance assessments of our stu-dents? (2) Can the quality and reliability be safely quantified? What methods should be used to quantify the reliability? (3) What effects would a data cleaning step bring up? Should we drop too unreliable assessments? Finally and more generally speaking we are interested in the question: What about the bad reputation of relevance assessment studies done with students or laymen?

The actual retrieval performance of the three value-added services and the pros and cons of each system is not the focus of this work. The services and their evaluation are a general framework for our studies on the inter-assessor reliability. In this paper we will analyse the quality of the assessments, measured in inter-assessor agreement by Fleiss' Kappa and Krippendorff's Alpha. While the first one is a standard measure recommended in Information Retrieval textbooks like Manning et al. [13], the second is rarely used in the IR and relevance assessment domain.

The paper is outlined as follows: We start with an overview on related work in the field of relevance assessment and the measurement of inter-assessor reliability in section 2. In section 3 we give a very short introduction of the three evaluated services, their implementation, materials and methods and the conducted relevance assessments. The results of our analysis are presented in section 4. We will close with a discussion and a look on future work in section 5.

## 2      Related Work

Information Retrieval (IR) test collections are typically built from a given set of documents, a set of topics and relevance judgments for documents that were made by a group of human assessors. The judgments are sometimes called assessments or ratings and therefore the people doing the judgments are called judges, assessors or raters. In this paper we will use the terms relevance assessments, assessors and inter-rater agreement respectively.

Since the early days of IR research and the construction of IR test collections a critical and general issue in conducting relevance assessment with more than one assessor per topic is the disagreement between the assessors. To get a feeling for the degree of agreement between the different assessors simple percentage-agreement or overlap counts were used in early TREC studies as documented by Voorhees [21], later Jaccard's coefficient (intersection/union) was used.

A variety of studies was compiled by Bailey et al. [4]. In their work they give a comprehensive overview on historical and recent studies on inter-assessor agreement and report on some characteristics of empirical studies of inter-assessor agreement in IR evaluation settings. We see a wide range of different settings from the number of relevance levels, number of topics, ratio of documents per topic, ratio of assessors per topic, to the kind of agreement measures that were reported in the original studies. For a short summary see table 1 which was compiled from the original paper from Bailey to allow a direct comparison to this study. In early years inter-assessor agreement measures like Jaccard coefficient or the intersection method were – and still today [16] are – used. These measures are getting unstable and unreliable as the number of categories or assessors increases. Later Kappa values from Cohen or Fleiss were used.

Fleiss' Kappa is a measure of inter-grader reliability or agreement for nominal or binary ratings and an extended version of Cohen's Kappa. While Cohen's Kappa is only suitable for two assessors, Fleiss' Kappa can be used for more than two assessors [7]. The computed Kappa values can be interpreted as the extent to which the observed amount of agreement among assessors exceeds what would be expected if all assessors made their ratings completely randomly. Kappa scores can range from -1 (no agreement) to 1 (full agreement). Landis and Koch [12] suggest interpreting the score as followed: $\kappa \leq 0$ = poor agreement, $0 \leq \kappa \leq 0.2$ = slight agreement, $0.2 \leq \kappa < 0.4$ = fair agreement, $0.4 \leq \kappa < 0.6$ = moderate agreement, $0.6 \leq \kappa < 0.8$ = substantial agreement, $0.8 \leq \kappa \leq 1$ = (almost) perfect agreement. These interpretations are not generally accepted and other interpretations are possible. Greve and Wentura [9] suggest interpreting scores $\kappa < .4$ as "not be taken too seriously" and values of $0.4 \leq \kappa < 0.6$ as acceptable. $0.75 \leq \kappa$ seems good up to excellent.

Besides Kappa some authors suggest to use Krippendorff's Alpha coefficient to measure agreement. While the use of Fleiss' and Cohen's Kappa is suggested in IR standard literature [13] and common practice in current research [2] Alpha coefficients are rather uncommon but are used in domains like opinion retrieval [5] or computational linguistics [3].

While for Kappa all assessors have to rate the same number of subjects and use the same scale the Alpha coefficient can usually handle more variations and computes reliabilities that are comparable across any numbers of assessors and values, different metrics, and unequal sample sizes. Krippendorff [11] argues for the use of Alpha in favour of other measure like Kappa because of its independence to the number of assessors and its robustness against imperfect data. For Krippendorff's Alpha there are the same doubts against such fixed and recommended values. Besides that Krippendorff himself pointed out that Alpha values are usually smaller than Kappa and that "except for perfect agreement, there are no magical numbers" [11]. Nevertheless he mentions $\alpha \geq 0.8$ as a threshold for perfect agreement.

**Table 1:** Overview on a compilation of studies (mainly taken from Bailey et al. [4] with additions from our own literature studies, marked by a citation) reporting characteristics of empirical studies of inter-assessor agreement.

| researchers | relev. levels | topics | docs/ topic | ass./ topic | agreement + measure |
|---|---|---|---|---|---|
| Lesk & Salton | 2 | 48 | 1268 | 2 | 31%, Jaccard |
| Cleverdon | 5 | 32 | 200 | 4 | - |
| Burgin | 3 | 100 | 1239 | 4 | 40-55%, Jaccard |
| Voorhees & Harman | 2 | 49 | 400 | 2 | 72%, overlay |
| Voorhees, Cormack | 2+3 | 49 | ≈124 | 2-5 | 33%, Jaccard |
| Sormunen | 4 | 38 | 31-200 | 2 | custom |
| Trotman et al. | 2 | 15 | 67-135 | 3-5 | custom |
| Bailey et al. [4] | 3 | 33 | 53-176 | 3 | Cohen's $\kappa$ |
| Piwowarski et al. [16] | 2-4 | 20 | - | 2 | 23-31%, Jaccard |
| Schaer (this study) | 2 | 10 | 40-50 | 2-13 | Fleiss' $\kappa$ and Krippendorff's $\alpha$ |

# 3 Materials and Methods

During the last years we conducted several relevance assessment evaluation series using three different retrieval services. The evaluation was done three times during the winter terms at University of Applied Sciences, Darmstadt and two times at Humboldt University, Berlin respectively.

## 3.1 Evaluated Services

Standard IR methods like TF*IDF are text-centric, which means they propose a text-based relevance ranking: These methods assign a weight to term t in document d which is influenced by different occurrences of t and d. While in general these methods work rather well especially in special domains like digital libraries and domain specific databases problems like the "language problem" and the need for alternative rankings become clear. We developed three science-model-driven methods that try to overcome these retrieval issues:

(1) Search Term Recommenders (STR), which are an approach to compensate the long known language problem in Information Retrieval. STRs are based on statistical co-word analysis and build associations between query terms and controlled terms (i.e. from a thesaurus). The co-word analysis implies a semantic association between the uncontrolled and the controlled terms. In our setup we use STRs for automatic query expansion where the original query of the user is enhanced with "semantically near" terms from a controlled vocabulary.

(2) Bradfordizing is an alternative mechanism to re-rank result lists according to core journals to bypass the problem of very large and unstructured result sets. The approach of Bradfordizing is to use characteristic concentration effects (Bradford's law of scattering) that appear typically in journal literature. Documents in core journals – journals that publish frequently on a topic – are ranked higher than documents that were published in journals from the following zones.

(3) Author centrality is another way of re-ranking result sets. Here the concept of centrality in a network of authors is an additional approach for the problem of large and unstructured result sets. The intention behind this ranking model is to make use of knowledge about the interaction and cooperation behaviour in special fields of research. The model is based on a network analytical view and differs greatly from conventional text-oriented ranking methods like TF*IDF.

## 3.2 Evaluation Setup

In our setup we used the SOLIS database with approx. 370.000 single documents from the social science domain. The database largely consists of metadata on scientific literature and is a superset of the GIRT corpus used in the TREC and CLEF evaluation campaigns. We only used a subset of the available metadata so that the assessed documents included title, abstract, author names and controlled keywords. We intentionally left out information like the publication year, publishers or the journal the documents were published in since we want our assessors to solely rely on the actual content information not additional hints that might let them draw conclusions from the currency or the reputation of a journal or publisher. The assessment system, which was built on top of the IRSA toolkit[1] and all documents were in German. All written examples in this paper are translated.

In our assessment each participants had to complete and assess one concrete search task, which was taken from the CLEF campaign. After a briefing each student had to choose one out of ten different predefined topics (namely CLEF topics 83, 84, 88, 93, 96, 105, 110, 153, 166 and 173). Topic title and the description were presented to form the information need (see table 1). Since the assessors were no domain experts in the social science domain we choose these topics because of their broad connection to youth, media, education, Germany in general and their ability to be used as common-sense retrieval tasks.

**Table 2:** Ten topics taken from the CLEF campaign, which were used in the relevance assessments.

| Topic | Title | Description |
|---|---|---|
| 83 | Media and War | Find documents on the commentatorship of the press and other media from war regions. |
| 84 | New Media in Education | Find documents reporting on benefits and risks of using new technology such as computers or the Internet in schools. |
| 88 | Sports in Nazi Germany | Find documents about the role of sports in the German Third Reich. |
| 93 | Burnout Syndrome | Find documents reporting on the burnout syndrome. |
| 96 | Costs of Vocational Education | Find documents reporting on the costs and benefits of vocational education. |
| 105 | Graduates and Labour Market | Find documents reporting on the job market for university graduates. |
| 110 | Suicide of Young People | Find documents investigating suicides in teenagers and young adults. |
| 153 | Childlessness in Germany | Information on the factors for childlessness in Germany |
| 166 | Poverty in Germany | Research papers and publications on poverty and homelessness in Germany. |
| 173 | Propensity towards violence among youths | Find reports, cases, empirical studies and analyses on the capacity of adolescents for violence. |

---

[1] http://sourceforge.net/projects/irsa/

The assessors saw a pooled list of result documents, so the origin of each document was disguised. The pool was formed out of the top n=10 ranked documents from each service and the initial TF*IDF ranked result set from the Solr search engine, respectively. Duplicates were removed, so that the size of the sample pools in 2010 was between 34 and 39 documents each. In 2011 and 2012 we added a so-called random ranking service. This service just randomly takes 10 documents from the original Solr query, which resulted in slightly larger result sets in 2011 and 2012. The assessors could choose to judge relevant or not relevant (binary decision).

### 3.3 Participants

A total of n=188 undergrad library and information science students contributed to the evaluation. They did a total of 9,226 single document assessments. Because some of the assessors didn't judge all of the documents we had to filter out some of the assessments. After a data cleaning step n=168 students remain in the data set. We discarded all assessments with more than 5% error rates (e.g. more than 2 documents missing from a theoretical data pool of 40 documents). In 2010 we had a total of 75 students doing the assessments, in 2011 we had 57 and in 2012 36 students participated. As stated above the evaluation was done three times during the winter terms at University of Applied Sciences, Darmstadt and two times at Humboldt University, Berlin respectively.

### 3.4 Computing Inter-assessor Agreement

We briefly list the basic approaches to compute Fleiss' Kappa and Krippendorff's Alpha to get a feeling for the two computations. In general both methods try to compute the amount of agreement by defining agreement as

$$\text{Agreement} = 1 - \frac{D_o}{D_e} = 1 - \frac{\text{Observed Disagreement}}{\text{Expected Disagreement}}$$

but they differ in the way they operationalize these computations (for a more comprehensive description see [10]).

Given a generic two by two contingency table 3 with the proportions a, b, c and d, where a + d is the observed agreement and b + c is the disagreement. The proportion of 0s in the data is given by $\bar{p} = (p_A + p_B)/2$ and the proportion of 1s by $\bar{q} = (q_A + q_B)/2$ or $1 - \bar{p}$. n is the number of 0s and 1s used jointly.

Kappa and Alpha are now computed by:

$$\kappa = 1 - \frac{b + c}{p_A q_b + p_B q_A}$$

and

$$\alpha = 1 - \frac{b + c}{\frac{n}{n-1} 2\bar{p}\bar{q}}$$

**Table 3:** Two by two contingency table (taken from [11])

|  | | Assessor A | | |
|---|---|---|---|---|
|  | | 0 | 1 | |
| Assessor B | 0 | a | b | $p_B$ |
|  | 1 | c | d | $q_B$ |
|  | | $p_A$ | $q_A$ | 1 |

All listed Kappa and Alpha values were computed using the R statistics software [17] and the irr package [8] respectively.

## 4 Results

We report on the outcomes of the inter-assessor agreements and on the implications these agreements or disagreements have on the evaluation of the initially described retrieval services when we drop the unreliably assessments from our data set.

### 4.1 Inter-assessor Agreement

The results of the inter-assessor agreement tests are listed in table 4. They are grouped per year and average values are given in the last columns and the last line. We can see that the average number of assessors per topic is between 4 and 8.7, the average Kappa values are between 0.210 and 0.524. Alpha values are generally below the Kappa values and the average Alpha values are between -0.018 and 0.279. Kappa values are all in the region of "fair" to "moderate" agreement but the Alpha values are far away from being "acceptable". The general agreement rate is low.

When we apply a Pearson correlation we get a relatively weak correlation coefficient of 0.447 on the average values. The highest correlation on a per year basis is the on from 2010 with 0.581, the other correlation coefficient are 0.406 for 2011 and 0.326 for 2012. Nevertheless we can see some essential misinterpretation on a per topic/year basis. While topic 96 in the year 2012 had one of the highest Alpha values the corresponding Kappa values is nearly 0. The opposite is true for topic 83 from 2010: here one of the highest Kappa values of 0.535 only got an Alpha value of 0.12.

We can see large differences between the different topics and years but the differences are (1) connected to the number of students and (2) the specific topic. While in 2010 the number of student assessors per topic was 7.5 and the correlation between Kappa and Alpha was 0.581, in 2012 only 3.6 students per topic had a lower correlation coefficient. The same is true to specific topics. Topics 153 and 173 both got very low Alpha and Kappa values although they were judged by 5 students on average.

**Table 4:** Inter-assessor agreement measured with Fleiss' Kappa and Krippendorff's Alpha for the years 2010 – 2012 and the corresponding average over all three years. The number of assessors per topic is given by n.

| Topic | 2010 | | | 2011 | | | 2012 | | | Average | | |
|---|---|---|---|---|---|---|---|---|---|---|---|---|
| | n | α | κ | n | α | κ | n | α | κ | n | α | κ |
| 83 | 13 | .120 | .535 | 8 | .229 | .412 | 5 | .092 | .318 | 8.7 | .147 | .421 |
| 84 | 9 | .165 | .283 | 5 | .073 | .480 | 3 | .169 | .366 | 5.7 | .136 | .376 |
| 88 | 6 | .181 | .528 | 3 | .327 | .257 | 5 | .197 | .550 | 4.7 | .235 | .445 |
| 93 | 10 | .036 | .330 | 5 | .375 | .713 | 3 | .195 | .529 | 6.0 | .202 | .524 |
| 96 | 2 | .293 | .591 | 9 | .186 | .113 | 4 | .358 | .001 | 5.0 | .279 | .235 |
| 105 | 4 | .125 | .536 | 4 | .068 | .345 | 4 | .052 | .307 | 4.0 | .082 | .396 |
| 110 | 5 | .148 | .223 | 8 | .104 | .386 | 4 | .308 | .413 | 5.7 | .187 | .341 |
| 153 | 9 | -.003 | .194 | 7 | .012 | .304 | 3 | -.063 | .132 | 6.3 | -.018 | .210 |
| 166 | 8 | .100 | .382 | 5 | .274 | .505 | 2 | .236 | .536 | 5.0 | .203 | .474 |
| 173 | 9 | .076 | .433 | 3 | .000 | .297 | 3 | -.081 | .084 | 5.0 | -.002 | .271 |
| avg. | 7.5 | .124 | .403 | 5.7 | .165 | .381 | 3.6 | .146 | .323 | 5.6 | .145 | .369 |

### 4.2 The Effects of Dropping Unreliable Assessments

Since the agreement rates measured by Kappa and Alpha reported in section 4.1 were below the recommended values for "acceptable" agreements we decided to measure the effects of data cleaning. Given the fact that there are no "magic numbers" we tried to pick thresholds that can be applied to the given data. If we had applied the high threshold reported in section 2 of κ, α ≥ 0.8 no single assessment would have remained in the data set.

In table 5 we see two different result sets containing the precision values (p@10) for the different services on a per topic basis. The first column set contains the unfiltered judgments from all assessors. Only the obviously wrong and sparse data sets were cleaned from this one (see section 3.3). The second column set contains the remaining results after all assessments with κ < 0.4 were removed from the result set. The same method is applied for the last column set where the threshold was α < 0.1.

Topics 153 (Childlessness in Germany) and 173 (Propensity towards violence among youths) contained the most inconsistencies. In all three years the Kappa and Alpha values were below the thresholds (only the Kappa values from 2010 were above the threshold). This way almost no assessments remained so that the two topics were mostly dropped for the Kappa-filter and completely dropped for the Alpha-filter. In total we had to drop 17 out of 30 assessment sets due to the Kappa filter and 11 due to the Alpha filter. For the Kappa filter no single topic had reliable assessments for all three years.

**Table 5:** Precision@10 values for five different retrieval services: SOLR (TF*IDF ranked, unprocessed baseline), RAND (the same baseline set but random ranking), AUTH (alternative ranking based on author centrality), BRAD (alternative ranking based on core journals, Bradfordizing) and STR (Query Expansion with controlled thesaurus terms). The left column set shows the unfiltered results from all assessors. The two right column sets are filtered with Fleiss Kappa and Krippendorff's Alpha, respectively. Empty cells are dropped values in all three years due to a too low inter-assessor agreement rate. The last line shows root mean square error between the unfiltered and filtered results.

| Topic | Original, unfiltered results (o) | | | | | Filtered with Kappa > .4 ($f_\kappa$) | | | | | Filtered with Alpha > .1 ($f_\alpha$) | | | | |
|---|---|---|---|---|---|---|---|---|---|---|---|---|---|---|---|
| | SOLR | RAND | AUTH | BRAD | STR | SOLR | RAND | AUTH | BRAD | STR | SOLR | RAND | AUTH | BRAD | STR |
| 83 | .75 | .39 | .47 | .27 | .75 | .74 | .30 | .43 | .22 | .74 | .74 | .30 | .43 | .22 | .74 |
| 84 | .77 | .35 | .32 | .64 | .57 | .79 | .31 | .30 | .65 | .51 | .80 | .43 | .30 | .61 | .54 |
| 88 | .47 | .45 | .14 | .66 | .54 | .47 | .54 | .16 | .69 | .49 | .47 | .42 | .13 | .66 | .54 |
| 93 | .68 | .46 | .68 | .73 | .57 | .63 | .44 | .62 | .71 | .41 | .63 | .44 | .62 | .71 | .41 |
| 96 | .40 | .45 | .80 | .59 | .49 | .40 | | .85 | .70 | .35 | .41 | .45 | .82 | .61 | .47 |
| 105 | .54 | .46 | .63 | .51 | .69 | .67 | | .65 | .59 | .45 | .67 | | .65 | .59 | .45 |
| 110 | .66 | .51 | .71 | .35 | .84 | .70 | .45 | .68 | .30 | .83 | .68 | .49 | .71 | .37 | .85 |
| 153 | .53 | .36 | .47 | .51 | .56 | | | | | | | | | | |
| 166 | .18 | .46 | .68 | .55 | .74 | .23 | .48 | .70 | .53 | .84 | .21 | .48 | .68 | .54 | .76 |
| 173 | .47 | .70 | .63 | .51 | .58 | .40 | | .58 | .49 | .74 | | | | | |
| avg. prec. | .55 | .46 | .55 | .53 | .63 | .56 | .42 | .55 | .54 | .60 | .57 | .43 | .54 | .54 | .60 |
| RMSerr(o,f) | | | | | | .03 | .05 | .06 | .05 | .12 | .02 | .03 | .05 | .05 | .10 |

To quantify the difference between the filtered and the unfiltered assessments sets and their values we applied the root mean square (RMS) error:

$$RMSerr(o,f) = \sqrt{\frac{1}{N}\sum_{i=1}^{N}(o_i - f_i)^2}$$

where $o_i$ and $f_i$ are the original/unfiltered and filtered values, respectively.

The RMSerr values are reported on the last line. We see moderate but considerable error rates between the unfiltered and filtered results. For the services SOLR, RAND, AUTH and BRAD the values are roughly around 0.05 for the Kappa-filtered and a little lower for the Alpha-filtered. The STR error values are 0.10 for the Kappa-filtered and 0.12 for the Alpha-filtered.

## 5   Discussion and Conclusion

When we look at the general agreement rate of our assessors we see a rather large range of results. In general and on average the agreement rates are fair to moderate but far away from being substantial or even perfect. On first sight the bad reputation of students doing relevance assessments seems legitimate: In the terms of Bailey et al. we were using "bronze standard judges" – so a perfect agreement could not be expected. On the other hand we see large differences between the different years and topics. Although we only had a small number of 10 topics in this study (which is quite small compared to the usual 25 – 50 topics suggested [20]) we had a high number of 5.6 assessors per topic. Since we saw that generally Kappa values are more prone to different numbers of assessors and does not scale that well compared to Alpha values we argue that beside the general practice of computing percentage overlaps, Jaccard coefficients and Fleiss' Kappa other values like Krippendorff's Alpha should be considered to get a more precise quantification of the agreement of the assessors and therefore a hint on the reliability of the collected assessment data.

Is there a general rule of thumb on how many assessors per query are necessary? Ideally, a large number of assessors per query should be used in an assessment. Recent approaches of using crowd-sourcing methods like Amazon's Mechanical Turk to do large-scale evaluations without domain experts are exactly going into that direction. But in the light of the rather low inter-assessor agreement rates in this controlled evaluation setup the uncontrolled situations in the crowd-sourcing approaches are debateable. So far and to our knowledge no Kappa or Alpha studies were done in this area yet. Studies by Alonso et al. [1] only reported on Jaccard coefficient and overlap counts.

What are good topics for lay assessors like students? Are their "easy" or "hard" topics in our assessment? Given the multidimensionality of relevance and the various relevance criteria users employ to judge the relevance, like described by Borlund [6], we should further analyse the observed disagreements and the connection to certain topics. We actually know very few on the motivation and the reasons for the disagreements in our assessment scenario.

When we apply the computed agreement rates to locate and filter out disputable assessment sets we see clear effects on the measured retrieval performance. In some cases this effect is quite drastic – like the different performance rates of the STR. This is in line with the general understanding of inter-assessor studies in other domains: "Reliability is […] a prerequisite for demonstrating the validity of the coding scheme – that is, to show that the coding scheme captures the 'truth' of the phenomenon being studied" [3]. But by computing the agreement or disagreement of assessors we can only draw conclusions on the *stability* and *reproducibility* of our data, not inevitably the *accuracy* of our results. To compute the last we would need a gold standard, which does not exist in our setting.

We see the effects of our filtering as a clear indicator that disagreement affects the reliability of evaluations. However "no consistent conclusion on how disagreement affects the reliability of evaluation has yet been drawn" [19] in the IR community. We should be carful drawing conclusions from unfiltered results (like "The STR approach clearly outperforms the other retrieval services."). We therefore suggest not to work with unfiltered results or – since thresholds are always debateable and can be interchanged to higher or lower values – to clearly document the immanent differences and disagreements between the assessors. The differences should be presented in the results using standard measures of inter-assessor agreement like Cohen's/Fleiss' Kappa or Krippendorff's Alpha. This way we would make a huge step towards more sound evaluation data sets.

**Acknowledgements.** We would like to thank Philipp Mayr (University of Applied Sciences, Darmstadt) and Vivien Petras (Humboldt University, Berlin) who supervised the students during the winter semesters of 2010-2012 and Hasan Bas and Peter Mutschke who developed the assessment component and the Author Centrality library respectively that were used in the evaluation systems. This work was partly funded by DFG (grant no. SU 647/5-2).

# References

1. Alonso, O., Schenkel, R., Theobald, M.: Crowdsourcing Assessments for XML Ranked Retrieval. In: Gurrin, C., He, Y., Kazai, G., Kruschwitz, U., Little, S., Roelleke, T., Rüger, S., and van Rijsbergen, K. (eds.) Advances in Information Retrieval. pp. 602–606 Springer Berlin / Heidelberg (2010).
2. Arguello, J., Diaz, F., Callan, J., Carterette, B.: A methodology for evaluating aggregated search results. Proceedings of the 33rd European conference on Advances in information retrieval. pp. 141–152 Springer-Verlag, Berlin, Heidelberg (2011).
3. Artstein, R., Poesio, M.: Inter-coder agreement for computational linguistics. Comput. Linguist. 34, 4, 555–596 (2008).
4. Bailey, P., Craswell, N., Soboroff, I., Thomas, P., de Vries, A.P., Yilmaz, E.: Relevance assessment: are judges exchangeable and does it matter. Proceedings of the 31st annual international ACM SIGIR conference on Research and development in information retrieval. pp. 667–674 ACM, New York, NY, USA (2008).
5. Bermingham, A., Smeaton, A.F.: A study of inter-annotator agreement for opinion retrieval. Proceedings of the 32nd international ACM SIGIR conference on Research and development in information retrieval. pp. 784–785 ACM, New York, NY, USA (2009).
6. Borlund, P.: The concept of relevance in IR. Journal of the American Society for Information Science and Technology. 54, 10, 913–925 (2003).
7. Fleiss, J.L.: Measuring nominal scale agreement among many raters. Psychological Bulletin. 76, 5, 378–382 (1971).
8. Gamer, M., Lemon, J., Puspendra Singh, I.F.: irr: Various Coefficients of Interrater Reliability and Agreement. (2010).
9. Greve, W., Wentura, D.: Wissenschaftliche Beobachtung : eine Einführung. Beltz, PsychologieVerlagsUnion, Weinheim (1997).

10. Krippendorff, K.: Computing Krippendorff's Alpha-Reliability, http://repository.upenn.edu/asc_papers/43, (2011).
11. Krippendorff, K.: Reliability in Content Analysis: Some Common Misconceptions and Recommendations. Human Communication Research. 30, 3, 411–433 (2004).
12. Landis, J.R., Koch, G.G.: The measurement of observer agreement for categorical data. Biometrics. 33, 1, 159–174 (1977).
13. Manning, C.D., Raghavan, P., Schütze, H.: Introduction to Information Retrieval. Cambridge University Press, Cambridge, UK (2008).
14. Mayr, P., Mutschke, P., Petras, V., Schaer, P., Sure, Y.: Applying Science Models for Search. 12. Internationales Symposium für Informationswissenschaft (ISI). (2011).
15. Mutschke, P., Mayr, P., Schaer, P., Sure, Y.: Science models as value-added services for scholarly information systems. Scientometrics. 89, 1, 349–364 (2011).
16. Piwowarski, B., Trotman, A., Lalmas, M.: Sound and complete relevance assessment for XML retrieval. ACM Trans. Inf. Syst. 27, 1, 1:1–1:37 (2008).
17. R Development Core Team: R: A Language and Environment for Statistical Computing. , Vienna, Austria (2011).
18. Schaer, P., Mayr, P., Mutschke, P.: Implications of Inter-Rater Agreement on a Student Information Retrieval Evaluation. In: Atzmüller, M., Benz, D., Hotho, A., and Stumme, G. (eds.) Proceedings of LWA2010 - Workshop-Woche: Lernen, Wissen & Adaptivitaet. , Kassel, Germany (2010).
19. Song, R., Guo, Q., Zhang, R., Xin, G., Wen, J.-R., Yu, Y., Hon, H.-W.: Select-the-Best-Ones: A new way to judge relative relevance. Inf. Process. Manage. 47, 1, 37–52 (2011).
20. Voorhees, E.M.: Topic set size redux. Proceedings of the 32nd international ACM SIGIR conference on Research and development in information retrieval. pp. 806–807 ACM, New York, NY, USA (2009).
21. Voorhees, E.M.: Variations in relevance judgments and the measurement of retrieval effectiveness. Inf. Process. Manage. 36, 5, 697–716 (2000).